\documentclass{PoS}

\newcommand{\nl}{n_{\ell}}

\newcommand{\mbar}{\overline{m}}
\newcommand{\MSR}{\mathrm{MSR}}

\newcommand{\MSb}{\overline{\mathrm{MS}}}
\newcommand{\df}{{\rm d}}

\title{Bottom and Charm Quark Mass Determination from Quarkonium at N$^3$LO}

\ShortTitle{Bottom and Charm Quark Mass Determination from Quarkonium at N$^3$LO}

\author{\speaker{Pablo G. Ortega}\\
        Departamento de F\'isica Fundamental and IUFFyM, Universidad de Salamanca, E-37008 
Salamanca, Spain\\
        E-mail\,: \email{pgortega@usal.es}}

\author{Vicent Mateu\\
       Departamento de F\'isica Fundamental and IUFFyM, Universidad de Salamanca, E-37008 
Salamanca, Spain\\Instituto de F\'isica Te\'orica UAM-CSIC, E-28049 Madrid, Spain\\
        E-mail\,: \email{vmateu@usal.es}}

\abstract{The non-perturbative nature of QCD at hadronic scales implied the 
development of phenomenological approaches such as 
quark models or, more recently, computer-based calculations using Lattice QCD. 
However, the unique properties of heavy quarkonium systems allow an entire 
calculation in terms of non-relativistic perturbative QCD.
In this work, the bottomonium spectrum, up to $n = 3$, and the ground state charmonium states,
are analyzed in the framework of \mbox{Non-Relativistic} Quantum Chromodynamics at N$^3$LO.
For bottomonium, finite charm quark mass effects in the QCD potential and the $\MSb$-pole
mass relation are incorporated to the highest known order, $\mathcal{O}(\varepsilon^3)$ in the
$\Upsilon$-scheme counting.
The bottom quark pole mass is expressed in terms of the MSR mass, a low-scale short-distance mass which
cancels the $u = 1/2$ renormalon of the static potential.
We study the $\nl = 3$ and $\nl = 4$ schemes, finding a negligible difference between the two
if finite $m_c$ effects are smoothly incorporated in the MSR mass definition.
We find that bottomonium $n=3$ states are not well behaved within perturbative NRQCD. Hence, 
fitting to the $n=1,2$ $b\bar b$ states we obtain $\mbar_b(\mbar_b) = 4.216\pm 0.039$ GeV.
Similarly, from the lowest lying charmonium states we find $\mbar_c(\mbar_c)=1.273 \pm 0.054$ GeV. 
}

\FullConference{XVII International Conference on Hadron Spectroscopy and Structure - Hadron2017\\
		25-29 September, 2017\\
		University of Salamanca, Salamanca, Spain}

\begin{document}

\vspace*{-.25cm}
\section{Introduction}
\vspace*{-.25cm}

%Traditionally, several phenomenological approaches has been developed to deal with the non-perturbative regime of QCD at hadronic scales, 
%such as quark models~\cite{Godfrey:1985xj,Segovia:2016xqb} or Lattice QCD~\cite{Mohler:2017ibi,Prelovsek:2013cta}.
%However, heavy quarkonium systems are unique, as their dynamics allow an alternative approach by means of non-relativistic QCD (NRQCD)
%using perturbation theory.
%
\noindent
Heavy quarkonium systems are unique, as their dynamics allow a full calculation using
non-relativistic QCD (NRQCD) with perturbative methods.
Within NRQCD the calculation of the energy levels of heavy quarkonium relies on the 
accurate description of the static QCD potential $V_{\rm QCD}(r)$.
Most recent calculations computed the energy levels of the lower-lying 
bottomonium states up to $\mathcal{O}(\alpha_s^5 m)$ and $\mathcal{O}(\alpha_s^5 
m \log \alpha_s)$ using pNRQCD~\cite{Brambilla:1999xf}, which describes the 
interactions of a non-relativistic system with ultrasoft gluons organizing the 
perturbative expansions in $\alpha_s$ and the velocity of heavy quarks 
systematically. A closed expression for arbitrary quantum numbers can be found 
in~\cite{Kiyo:2014uca}.

The convergence  of the perturbative expansion depends, though, on the 
short-distance mass scheme employed to ensure the $\mathcal{O}(\Lambda_{\rm 
QCD})$ renormalon cancellation. The authors of Ref.~\cite{Brambilla:2001fw} 
employed the well-known $\overline{\rm MS}$ scheme, commonly used for physical 
situations in which the relevant scale is of the order or larger than the heavy 
quark mass. For heavy quarkonium the typical scale is much smaller, therefore the 
results can be substantially improved by switching to a low-scale short-distance 
scheme.

In this work we study the predictions for the energy levels of 
heavy quarkonium at N$^3$LO using the MSR scheme~\cite{Hoang:2017suc}, and 
determine the bottom quark mass, including finite charm quark mass effects, using states with $n\le 3$ and
the charm quark mass from the low-lying charmonium spectrum.
A careful study of scale variation is performed, using the idea that the argument of perturbative logarithms should roughly vary between $1/2$ and $2$,
while keeping at all times the renormalization scale above $1\,$GeV such that perturbation theory is still valid.
A more detailed description of this work has been recently presented in Ref.~\cite{Mateu:2017hlz}. The reader is kindly referred
to the latter work for further details.

\vspace*{-.25cm}
\section{Computation of $Q\overline Q$ Bound States}
\vspace*{-.25cm}

%%%%%%%%%%%%%%%%%%%%%%%%%%%%%%%%%%%%%%%%%%%%%%%%%%%%%%%%%%%%%%%%%%%%%%%%%%%%%%%%%%%%%%%%%%%%%%%%%%%%%%%%%%%%%%%%%%%%%%%%%%%%%%
%%%%%%%%%%%%%%%%%%%%%%%%%%%%%%%% BOTTOM PAPER %%%%%%%%%%%%%%%%%%%%%%%%%%%%%%%%%%%%%%%%%%%%%%%%%%%%%%%%%%%%%%%%%%%%%%%%%%%%%%%%
\noindent
The energy of a  non-relativistic $Q\overline Q$ bound state with arbitrary quantum numbers and with $\nl$ massless active flavors 
reads, in the pole scheme reads~\cite{Penin:2002zv,Beneke:2005hg,Kiyo:2014uca}\,:
\begin{equation}\label{eq:EXpole}
 E_X(\mu,\nl) = 2\,m_Q^{\rm pole}\!
\Bigg[1-\frac{C_F^2\,\alpha^{(\nl)}_s(\mu)^2}{8n^2}\sum_{i=0}^{\infty}\bigg(\frac{\alpha^{(\nl)}_s(\mu)}{4\pi}\bigg)^{\!\!i}\,
\varepsilon^{i+1}P_i(L_{\nl})\Bigg],
\end{equation}
with
\begin{equation}
L_{\nl}=\log\!\bigg(\frac{n\mu}{C_F\alpha_s^{(\nl)}(\mu)m_Q^{\rm pole}}\bigg)+ H_{n+\ell}\,,\qquad
 P_i(L_{\nl}) = \sum_{j=0}^i\,c_{i,j}\,L_{\nl}^j\,,\nonumber
\end{equation}
where $H_{n}$ is the harmonic number and $\varepsilon$ is a bookkeeping parameter used to
organize the various orders in the $\Upsilon$-expansion scheme.
Imposing $\mu$ independence of the energy levels, the $c_{i,j}$ coefficients can be recursively calculated 
from $c_{i,0}$, which have been calculated up to $i=3$ in Refs.~\cite{Brambilla:2001qk,Kiyo:2013aea}.

A precise determination of the heavy quark masses directly from Eq.~(\ref{eq:EXpole}) is inadequate, as such equation inherits the
 $u=1/2$ renormalon of the static potential. Thus, the pole mass must be expressed in terms
of a short-distance mass. Furthermore, a low-scale short-distance mass is advisable to avoid large logarithms of the ratio of the non-relativistic scale and the quark
mass. In this work, two equivalent versions of the MSR mass~\cite{Hoang:2008yj,Hoang:2017suc} will be used. 

The $\MSR$ scheme was built as a natural extension of the $\MSb$ mass for renormalization scales below the heavy quark mass.
Its definition is derived directly from the $\MSb$-pole mass relation and, in contrast with the $\MSb$ mass, which has only logarithmic
dependence on the scale $\mu$, $\MSR$ mass has logarithmic and linear dependence on an infrared scale R\,:
\begin{equation}\label{eq:MSRdef}
  \delta m_Q^\MSR \equiv m_Q^{\rm pole}-m_Q^\MSR(R) = R\sum_{n=1}^\infty a_n(\nl)\left(\frac{\alpha^{(\nl)}_s(R)}{4\pi}\right)^{\!\!n}\,.
\end{equation}
The $a_{n}$ are derived from the $\MSb$ pole mass relation, and are different for the \emph{practical} and \emph{natural} versions of the
MSR mass, which are two alternative but equivalent ways to change from a scheme with $(n_\ell+1)$ dynamical flavors 
to another with only $n_\ell$. On the one hand, the 
\emph{practical $\MSR$ mass} (MSRp) uses the threshold matching relation of the strong coupling to rewrite $\alpha_s^{(n_\ell+1)}(\overline{m})$ in terms of $\alpha_s^{(n_\ell)}(\overline{m})$ and,
on the other hand, the \emph{natural $\MSR$ mass} (MSRn) directly integrates out the heavy quark Q from the $\MSb$ to pole relation setting all diagrams with heavy quark loops to zero.

The R dependence of the $\MSR$ mass is described by\,:
\begin{equation}\label{eq:Revol}
-\,\frac{\df}{\df R}m_Q^{\MSR}(R)=\gamma^R[\alpha_s^{(\nl)}(R)]=\sum_{n=0}^\infty \gamma_n^R \left(\frac{\alpha_s^{(\nl)}(R)}{4\pi}\right)^{\!\!n+1},
\end{equation}
where $\gamma_n^R = a_{n+1}-2\sum_{j=0}^{n-1} (n-j)\,\beta_j\, a_{n-j}$ are the R-anomalous dimension coefficients~\cite{Hoang:2017suc}.

Effects from the finite charm quark mass contribute to the relation of the pole mass to the $\MSR$ mass
and to Eq.~(\ref{eq:EXpole}).
Corrections to the binding energy result in an energy-shift to the heavy quarkonium mass and have been calculated up to 
$\mathcal{O}(\varepsilon^3)$~\cite{Eiras:2000rh,Hoang:2000fm,Beneke:2014pta}.
In the pole mass scheme they can be written as\,:
\begin{equation}\label{eq:EXpoleCorr}
E_X(\mu,\nl,m_Q^{\rm pole},m_q^{\rm pole}) = E_X(\mu,\nl,m_Q^{\rm pole})+\varepsilon^2\delta E_X^{(1)}+\varepsilon^3\delta E_X^{(2)}\,.
\end{equation}
The exact $\delta E_X^{(2)}$ term has only been computed for the $\ell=0$ bottomonium states~\cite{Hoang:2000fm,Beneke:2014pta}.
For the rest of states we follow the approach used in Ref.~\cite{Brambilla:2001qk}, where the authors employ the $m_c\to \infty$ limit,
which we complete by requiring that the correction in the $\nl$ scheme vanishes for $m_c\to 0$.

The corrections from the finite charm quark mass to the $\MSR$-pole mass relation start at $\mathcal{O}(\alpha_s^2)$\,:
\begin{eqnarray}\label{eq:MSRdefMC}
&\delta m_Q^\MSR(R,\overline{m}_c) = \delta m_Q^\MSR(R) + R\,\ \sum_{k=2}\Delta_{\mbar_c}^{(k)}(\xi)
\left(\frac{\alpha_s^{(n_\ell)}(R)}{4\pi}\right)^{\!\!k},
\end{eqnarray}
where $\xi=\mbar_c/R$ and $\delta m_Q^{\rm MSR}$ is given in Eq.~(\ref{eq:MSRdef}).
The $\varepsilon^2$ correction term has been calculated exactly in Ref.~\cite{Gray:1990yh}. 
The $\varepsilon^3$ correction was analytically calculated in Ref.~\cite{Bekavac:2007tk}, but due to its complexity
we have used a numerical approximation using a Pad\`e parametrization, which is accurate up to 8
significant digits in the range $0\le \xi \le 1$, enough for our purposes~\cite{Mateu:2017hlz}.

The aforementioned finite charm mass corrections also modify the R-anomalous dimension\,:
\begin{equation}\label{eq:RevolMC}
-\,\frac{\rm d}{{\rm d}R}m_Q^{\MSR}(R) \,=\, \gamma^R[\alpha_s^{(\nl)}(R)] + \sum_{n=1}\delta\gamma_n^R(\xi) \left(\frac{\alpha_s^{(\nl)}(R)}{4\pi}\right)^{\!\!n+1},
\end{equation}
with \,\,$\delta\gamma_n^R(\xi)  = \Delta_n(\xi)-\xi\,\frac{\df \Delta_n(\xi)}{\df \xi}-2\sum_{j=0}^{n-2} (n-j)\,\beta_j\, \Delta_{n-j}(\xi)$.

The $\mathcal{O}(\alpha_s^4)$ finite charm quark mass corrections are unknown so in our calculations 
we will cut the propagation of $\Delta_{\mbar_c}^{(2,3)}(\xi)$ terms to the $\mathcal{O}(\varepsilon^4)$ order, in order to
avoid possible cancellations among crossed terms.

Up to now we have explicitly account for the charm mass corrections in the $\nl$ scheme.
Alternatively, one can integrate out the charm quark, as
it is sufficiently large compared to typical NRQCD scales to assume it is near the decoupling limit ($\mbar_c\to\infty$). Therefore
we will study, together with the $\nl$ scheme with charm quark mass corrections, a scheme with $(\nl\!-\!1)$ dynamical flavors.

\vspace*{-.25cm}
\section{Results}
\vspace*{-.25cm}

%Fitting
\noindent
Our aim is to extract the $\MSb$ bottom and charm masses from the perturbative expression for the mass of heavy quarkonium states, contained
in Eq.~(\ref{eq:EXpole}) in the MSR mass scheme, fitting it against experimental values from the PDG.
For a given dataset, $\mbar_Q$ is obtained from the minimum of the following $\chi^2$ function, which depends on
a set of pairs of renormalization scales $\{\mu_n,R_n\}$, where the index $n$ runs over each principal quantum number in the dataset\,:
\begin{eqnarray}
\chi^2(\{\mu_n,R_n\})=\sum_i\left(\frac{M_i^{\rm exp}-M_i^{\rm pert}(\mu_i,R_i,\mbar_Q)}{\sigma_i^{\rm exp}}\right)^2.
\end{eqnarray}
The previous sum extends to the individual heavy quarkonium states in the dataset and $M_i^{\rm exp}$ and $\sigma_i^{\rm exp}$ are the experimental
masses and errors extracted from the PDG~\cite{Patrignani:2016xqp}.
Such fit would give us a best-fit $\MSb$ mass value as a function of the $\{\mu_n,R_n\}$ pair. This approach is taken because
the theoretical uncertainties are highly correlated among various states and the so-called d'Agostini bias~\cite{DAgostini:1993arp} emerges, 
which states that the global best-fit value is considerably lower than individual determinations from each state in the set.
An alternative way is to perform a weighted average of the best-fit results. Both methods are in quite good agreement, though this last option
leads to slightly larger perturbative uncertainties.

The renormalization scales $\mu$ and $R$ are varied independently, within reasonable ranges, to estimate the the perturbative uncertainties.
Such ranges depend on the convergence behavior for each heavy quarkonium state. 
Our procedure is to select ranges where the argument of the perturbative logarithms are roughly unity.
This implies that, for $n=1$ and $n=3$ states, the renormalization scales should be varied between 
$(\mu_1,R_1)\in[\,1.5,4\,]$\,GeV, whereas for states with $n=2$ the range \mbox{$(\mu_2,R_2)\in[\,1,4\,]$\,GeV} is preferred.
For charmonium we use the range \mbox{$1.2\,{\rm GeV}\ge\mu_{\rm charm}\ge 4\,{\rm GeV}$}, which gives a pattern very similar to the $n=1$ bottomonium states.
It is worth noticing that the $\mu_n $ and $R_n$ scales have to be varied together to account for theoretical correlations.

From the $\chi^2$ minimization we obtain the central value of $\mbar_Q$, calculated as the average of all the best-fits in the ($\mu,R$) grid, 
and the uncertainty coming from the experimental error of the heavy quarkonium masses, denoted as $\Delta^{\rm exp}$, also taken as the average in the grid.

The largest source of error is the one associated to the variation of the scales, that is, $\Delta^{\rm pert}$.
Its value is calculated as the difference of the maximum and minimum best-fit values in the grid.
Finally, there are additional errors originating from the uncertainty on the strong coupling $\Delta^{\alpha_s}$ %($\alpha_s^{(\nf=5)}(m_Z)=0.1181\pm0.0011$~\cite{Patrignani:2016xqp})
and, for the bottom (charm) mass fits, the charm (bottom) mass $\Delta^{\mbar_c}$ ($\Delta^{\mbar_b}$).

%%Results for bottom

\begin{figure*}[t]
	\center
    {\label{fig:Bottom-mass-states}\includegraphics[width=0.5\textwidth]{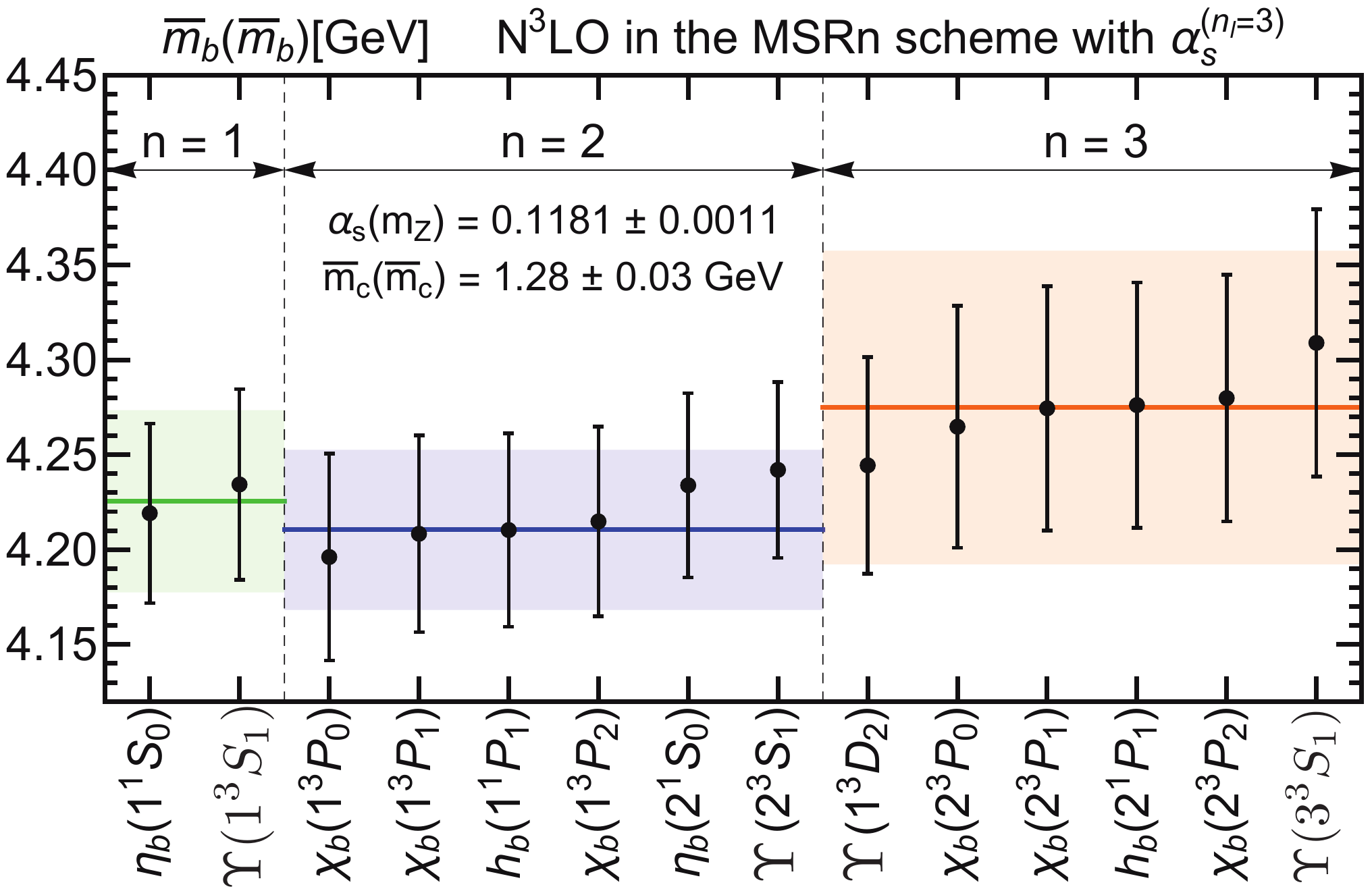}~~~}
    {\label{fig:Charm-mass-states}\includegraphics[width=0.47\textwidth]{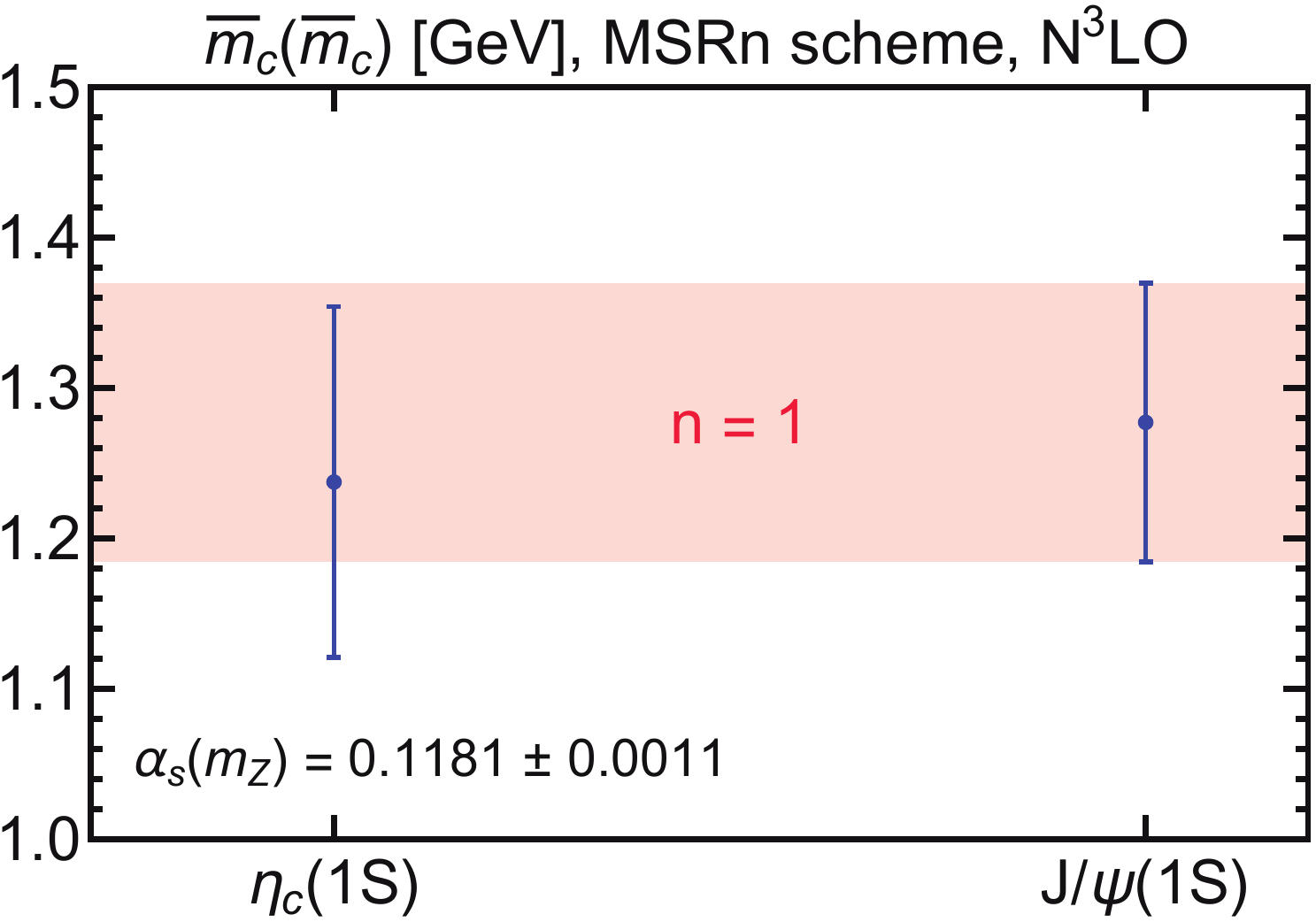}~~~}

 \caption{\label{fig:bottomResults} Left panel\,: Bottom quark mass determinations from individual fits to bottomonium states
with principal quantum number $n\le3$ (black dots with error bars) and global fits to $n = 1$ (green band), $n = 2$
(blue band), and $n = 3$ states (orange band). All computations at N$^3$LO in the MSRn scheme with $\nl = 3$ active flavors.
Right panel\,: Same for charm quark mass. Both panels\,: only perturbative uncertainties shown. }
\end{figure*}

We have carried out fits for $14$ individual bottomonium states (all states with $n\le3$) and global fits for sets with $n=1$ states, $n=2$ states, $n=3$ states, and 
different combinations such as all states with $n\le 2$ and $n\le 3$. Our results are shown in Fig.~\ref{fig:Bottom-mass-states} for the MSRn scheme with $\nl=3$.

The expansion with $\nl=3$ flavors is favored over $\nl=4$, as suggested by Ref.~\cite{Ayala:2014yxa}. 
We consider the set with ${n\le 2}$ states as our default, as such states can be accurately described within perturbation theory.
We take the MSRn scheme (which is theoretically cleaner) as our default, finding our final result for the bottom mass\,:
\begin{eqnarray}\label{eq:bottomFinal}
\!\!\!\!\!\!\!\mbar_b(\mbar_b) & = 4.216 \pm 0.009_{\rm exp} \pm 0.034_{\rm pert}\pm 0.017_{\alpha_s}\pm 0.0008_{\,\mbar_c}\,{\rm GeV}= 4.216 \pm 0.039\,{\rm GeV}.
\end{eqnarray}
The perturbative uncertainty clearly dominates over the rest, followed by the error from the strong coupling. The error due to the uncertainty on
the charm mass is negligible, as the experimental one.

%%%%%%%%%%%%%%%%%%%%%%%%%%%%%%%%%%%%%%%%%%%%%%%%%%%%%%%%%%%%%%%%%%%%%%%%%%%%%%%%%%%%%%%%%%%%%%%%%%%%%%%%%%%%%%%%%%%%%%%%%%%%%%%%%%%%%%%%%%%%%%%%%%%%%%%%%
%%%%%%%%%%%%%%%%%%%%%%%%%%%%%%%%%%%%%%%%%%%%%%%%%%%%%%%%%%%%%%%%%%%%%%%%%%%%%%%%%%%%%%%%%%%%%%%%%%%%%%%%%%%%%%%%%%%%%%%%%%%%%%%%%%%%%%%%%%%%%%%%%%%%%%%%%

%%Results for charm

Equivalently, the charm mass can be obtained from $c\bar c$ bound state calculations via Eq.~(\ref{eq:EXpole}).
%In this case there is no need to include lighter-quark corrections.
We will restrict ourselves to $n=1$ states and vary 
$\mu$ and $R$ between $1.2\,$GeV and $4\,$GeV, which yields order-by-order
convergence with moderate perturbative uncertainties. We will perform individual and global 
fits (for $n=1$), which are summarized at N$^3$LO in Fig.~\ref{fig:Charm-mass-states}. 
We find that the result of the global fit is almost identical to the individual fit to the
$J/\psi$ state, due to its extremely precise mass value with respect to the $\eta_c$ state.

As previously, we take the MSRn scheme as our default, so our final value for the charm quark mass from the global fit of $n=1$ states is\,:
\begin{eqnarray}\label{eq:charmFinal}
\!\!\!\!\!\!\!\!\!\!\!\mbar_c(\mbar_c) & = 1.273 \pm 0.0005_{\rm exp} \pm 0.054_{\rm pert}\pm 0.006_{\alpha_s}\pm 0.0001_{\mbar_b}\,{\rm GeV}= 1.273 \pm 0.054\,{\rm GeV}.
\end{eqnarray}
%
%For the error coming from $\mbar_b$ we took the 2016 world average \mbox{$\mbar_b = 4.18_{-0.03}^{+0.04}\,$GeV}~\cite{Patrignani:2016xqp}.
Again, we observe that theoretical uncertainties greatly dominate over $\alpha_s$ uncertainties,
and again experimental uncertainties (coming form the fit) are negligibly small. The $\mbar_b$ uncertainty is also negligible, 
as the dependence on the bottom  mass comes only from the $\alpha_s$ threshold matching from $5$ to $4$ flavors.
A more complete analysis of this work is to be found in Ref.~\cite{Mateu:2017hlz}.

%Conclusions

%%%%%%%%%%%%%%%%%%%%%%%%%%%%%%%%%%%%%%%%%%%%%%%%%%%%%%%%%%%%%%%%%%%%%%%%%%%%%%%%%%%%%%%%%%%%%%%%%%%%%%%%%%%%%%%%%%%%%%%%%%%%%%%%%%%%%%%%%%%%%%%%%%%%%%%%%
%%%%%%%%%%%%%%%%%%%%%%%%%%%%%%%%%%%%%%%%%%%%%%%%%%%%%%%%%%%%%%%%%%%%%%%%%%%%%%%%%%%%%%%%%%%%%%%%%%%%%%%%%%%%%%%%%%%%%%%%%%%%%%%%%%%%%%%%%%%%%%%%%%%%%%%%%

%\section*{Acknowledgments}

\emph{Acknowledgments}. 
This work has been partially funded by the Spanish MINECO ``Ram\'on y  Cajal'' program (RYC-2014-16022), the MECD grant FPA2016-78645-P, and 
the IFT ``Centro de Excelencia Severo Ochoa'' Program under Grant SEV-2012-0249; and by Junta de Castilla y Le\'on and European Regional Development Funds (ERDF) under
Contract no. SA041U16. 

\bibliographystyle{JHEP}

\bibliography{NRQCD}

\end{document}